\documentclass[pra,twocolumn,numerical,superscriptaddress]{revtex4-1}
\usepackage[latin9]{inputenc}
\usepackage[a4paper]{geometry}
\usepackage{color}
\usepackage{amsmath}
\usepackage{amssymb}
\usepackage{graphicx}

\makeatletter
\usepackage{dcolumn}
\usepackage{bm}
\usepackage{epsfig}
\usepackage{braket}

\renewcommand{\fnum@figure}{FIG.~\thefigure}
\usepackage{upgreek}

\makeatother

\begin{document}
\title{Laser induced persistent orientation of chiral molecules}
\author{Ilia Tutunnikov}
\thanks{I. T. and J. F. contributed equally to this work.}
\affiliation{AMOS and Department of Chemical and Biological Physics, Weizmann Institute of Science, Rehovot 7610001, Israel}
\author{Johannes Floß}
\thanks{I. T. and J. F. contributed equally to this work.}
\affiliation{Chemical Physics Theory Group, Department of Chemistry, and Center for Quantum Information and Quantum Control, University of Toronto, Toronto, Ontario M5S 3H6, Canada}
\author{Erez Gershnabel}
\affiliation{AMOS and Department of Chemical and Biological Physics, Weizmann Institute
of Science, Rehovot 7610001, Israel}
\author{Paul Brumer}
\thanks{paul.brumer@utoronto.ca}
\affiliation{Chemical Physics Theory Group, Department of Chemistry, and Center
for Quantum Information and Quantum Control, University of Toronto,
Toronto, Ontario M5S 3H6, Canada}
\author{Ilya Sh. Averbukh}
\thanks{ilya.averbukh@weizmann.ac.il}
\affiliation{AMOS and Department of Chemical and Biological Physics, Weizmann Institute
of Science, Rehovot 7610001, Israel}
\begin{abstract}
We show, both classically and quantum mechanically, enantioselective
orientation of gas phase chiral molecules excited by laser fields
with twisted polarization. Counterintuitively, the induced orientation,
whose direction is laser controllable, does not disappear after the
excitation, but stays approximately constant long after the end of
the laser pulses, behavior unique to chiral systems. We computationally
demonstrate this long-lasting orientation using propylene oxide molecules
(${\rm CH_{3}CHCH_{2}O}$, or PPO) as an example, and consider two
kinds of fields with twisted polarization: a pair of delayed cross-polarized
pulses, and an optical centrifuge. This novel chiral effect opens
new avenues for detecting molecular chirality, measuring enantiomeric
excess and separating enantiomers with the help of inhomogeneous external
fields.
\end{abstract}
\maketitle

\section{Introduction}

Chiral molecules exist in two enantiomeric forms. The two enantiomers
are mirror images of each other, and they are nonsuperimposable by
translation and rotation \citep{Cotton}. Molecular chirality is an
omnipresent natural phenomenon of extreme importance in physics, chemistry
and biology \citep{UniversalChirality}. The ability to discriminate
and separate mixtures of enantiomers is important, for example, in
drug synthesis as different enantiomers of chiral drugs may exhibit
strikingly different biological activity.

The related studies of gas phase chiral molecules focus on the measurements
of enantiomeric excess, handedness of a given compound, and on devising
techniques for manipulating mixtures containing both enantiomers \citep{ShapiroBrumer,Brumer2002,Lux2012,Patterson2013,Pitzer2013,Herwig2013,Lehmann2013,Janssen2014,Patterson2014,Steinbacher2015,Lux2015,Christensen2015,Kastner2016,Beaulieu2018,Pitzer2018,Fehre2019,Neufeld2019,Rozen2019,Leibscher2019}.
In addition, over the years more ambitious directions were considered
theoretically - laser assisted asymmetric synthesis, enantiomeric
interconversion and purification (e.g. \citep{Shapiro2000,ShapiroBrumer}
and references therein).

Recently, a pair of non-resonant delayed cross-polarized laser pulses
was proposed as a new tool for discrimination of chiral molecules
\citep{Yachmenev2016,Gershnabel2018,Tutunnikov2018} and the underlying
classical enantioselective molecular orientation mechanism was exposed
\citep{Gershnabel2018,Tutunnikov2018}. The approach was extended
to general fields with time-dependent polarization twisting in a plane.
Optical fields with fixed linear polarization are unable to induce
molecular orientation because of the symmetry of light interaction
with the induced dipole. Polarization twisting in a certain plane
breaks that symmetry and defines a preferred spatial direction perpendicular
to that plane that depends on the sense of polarization rotation.
When interacting with molecules, the twisted field has a two-fold
effect. First, it induces unidirectional rotation (UDR) of the most
polarizable molecular axis in the plane of polarization twisting \citep{Fleischer2009,Kitano2009,Khodorkovsky2011,Korech2013,Mizuse2015,Lin2015},
thereby orienting the averaged angular momentum vector $\braket{\boldsymbol{\ell}}$
perpendicular to the plane. In addition, in the case of chiral molecules,
the twisted field induces an orienting torque \textit{along} the most
polarizable axis, thus orienting the molecule itself perpendicular
to the above plane \citep{Gershnabel2018,Tutunnikov2018}. The direction
of orientation depends on both the sense of polarization twisting
and the handedness of the molecule. A pair of delayed cross-polarized
laser pulses \citep{Fleischer2009,Kitano2009} provides the simplest
example of the field with twisted polarization, but there are also
more complex fields, such as chiral pulse trains \citep{Zhdanovich2011,Bloomquist2012,Floss2012},
polarization-shaped pulses \citep{Karras2015,Prost2017,Prost2018}
and optical centrifuge \citep{Karczmarek1999,Villeneuve2000,Yuan2011,Korobenko2014,Korobenko2018}.
Most recently, the orientation of chiral propylene oxide (PPO) molecules
by means of an optical centrifuge was experimentally achieved in \citep{Milner2019},
thus providing the first demonstration of enantioselective laser control
over molecular rotation.

In this paper, we substantiate the enantioselective molecular orientation
by twisted fields in two significant ways. First, we provide a fully
quantum treatment of this behavior and, second, we show that the orientation
in chiral molecules is long-lived. Fully quantum studies on laser-driven
PPO are provided and compared to classical results. Two implementations
of the laser fields with twisted polarization are considered - a pair
of delayed cross-polarized pulses, and an optical centrifuge. The
laser-induced orientation is shown to be robust against normally detrimental
temperature effects.

The paper is organized as follows. In Section II, the conditions needed
for long-lasting field-free orientation in a pulse-excited molecular
ensemble are discussed, and we show that chiral molecules excited
by a laser field with twisted polarization satisfy these conditions.
In Section III, the results of classical and fully quantum simulations
of the enantio-selective orientation at thermal conditions are provided
and compared with one another. Section IV concludes the paper.

These results substantiate, quantum mechanically, the classical expectation
\citep{Gershnabel2018,Tutunnikov2018} that the orientation does not
disappear after the excitation, but stays at an approximately constant
level long after the end of the laser field.

\section{Classical Analysis \label{sec:Classical-Analysis}}

In this section we use classical mechanics to analyze conditions leading
to the existence of permanent orientation in a gas of field-free rotating
asymmetric-top molecules. We show that excitation of an isotropic
ensemble of chiral molecules by twisted fields satisfies these conditions.\\

\begin{figure}
\begin{centering}
\includegraphics[width=86mm]{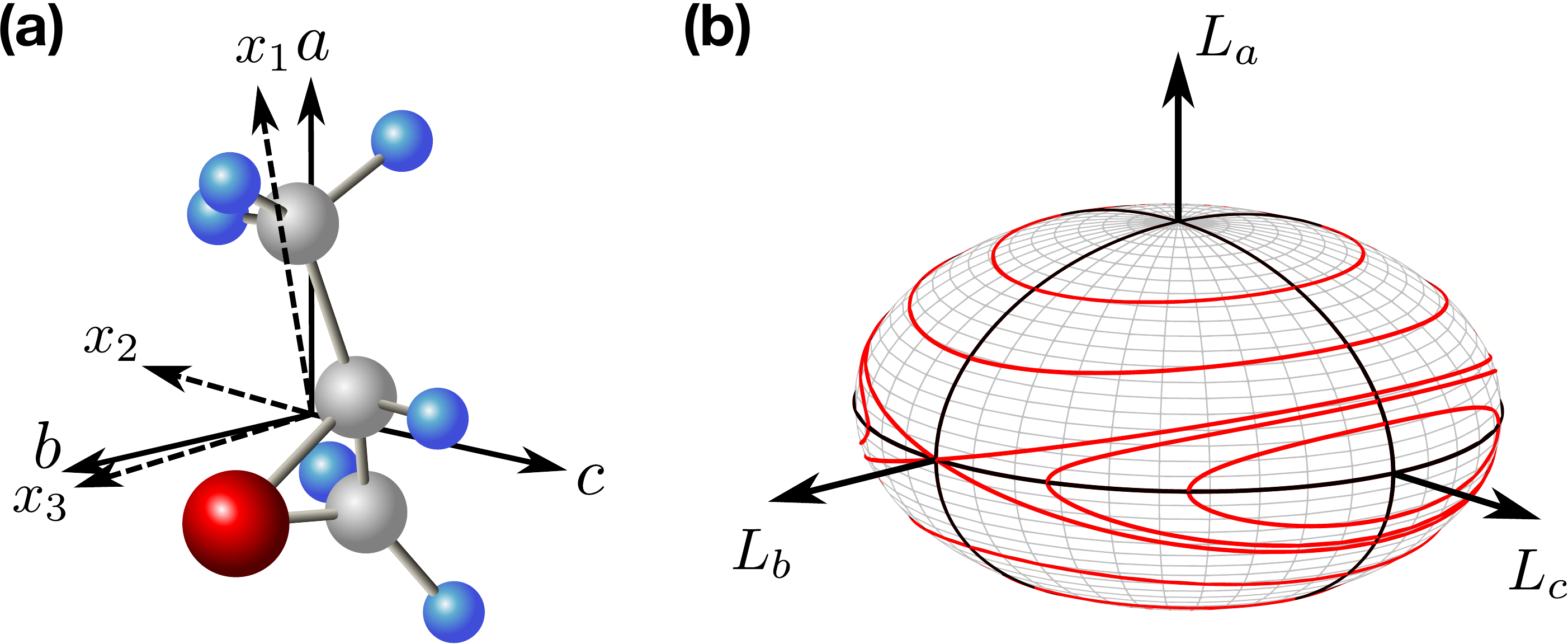}
\par\end{centering}
\caption{\textbf{(a) }(\emph{R})-Propylene oxide molecule. Atoms are color-coded:
gray - carbon, red - hydrogen, red - oxygen. Principal axes of inertia
tensor are shown as solid arrows and denoted by $a$, $b$ and $c$
$\left(I_{a}<I_{b}<I_{c}\right)$, while those of polarizability tensor
are shown as dashed arrows and denoted by $x_{1}$, $x_{2}$ and $x_{3}$
$\left(\alpha_{33}<\alpha_{22}<\alpha_{11}\right)$. \textbf{(b)}
Binet ellipsoid. In red - the allowed trajectories of the angular
momentum vector. \label{fig:Molecule-Binet}}
\end{figure}

\noindent \textbf{Background.} The Binet construction \citep{Goldstein,LandauLifshitzMechanics}
allows one to classify the trajectories traversed by the angular momentum
vector in the molecule-fixed frame defined by the three principal
axes of moment of inertia tensor. Figure \ref{fig:Molecule-Binet}(a)
shows one of the enantiomers (right handed, (\emph{R}))of our example
molecule, propylene oxide (${\rm CH_{3}CHCH_{2}O}$, or PPO) with
two sets of axes: principal axes of moment of inertia and polarizability
tensors. Hereafter, the letters $a$, $b$, $c$ refer to the inertia
principal frame, whereas the numbers 1, 2, 3 refer to the polarizability
principal frame.
We briefly describe the construction. Free motion of an asymmetric-top
rotor has four constants of motion, energy and the three components
of angular momentum expressed in an inertial frame (laboratory fixed
frame). In the frame of principal axes of the inertia tensor {[}see
Fig. \ref{fig:Molecule-Binet}(a){]}, the allowed trajectories of
the angular momentum vector satisfy the following equations
\[
\frac{L_{a}^{2}}{2EI_{a}}+\frac{L_{b}^{2}}{2EI_{b}}+\frac{L_{c}^{2}}{2EI_{c}}=1;\qquad\frac{L_{a}^{2}+L_{b}^{2}+L_{c}^{2}}{L^{2}}=1,
\]

\noindent where $I_{j}$ are the moments of inertia, $L_{j}$ are
the components of the angular momentum vector $\left(j=a,b,c\right)$,
$E$ is the rotational energy and $L$ is the magnitude of the angular
momentum vector. The moments of inertia are ordered according to $I_{a}<I_{b}<I_{c}$.
The first expression defines an ellipsoid with semi-axes $\sqrt{2EI_{a}}$,
$\sqrt{2EI_{b}}$ and $\sqrt{2EI_{c}}$; these coincide with the principal
axes of inertia tensor. The second expression defines a spherical
shell with radius $L$. The angular momentum vector tip moves on the
ellipsoid-spherical shell intersections. In addition to the six stationary
rotations about each of the three inertia tensor principal axes, the
allowed trajectories can be divided into sets of closed curves, as
shown in Fig. \ref{fig:Molecule-Binet}(b).

For $\sqrt{2EI_{a}}<L<\sqrt{2EI_{b}}$ and $\sqrt{2EI_{b}}<L<\sqrt{2EI_{c}}$,
the trajectories are divided into two sets of curves $T_{k}$ ($k=a,c$)
enclosing the poles on $a$ and $c$ axes, respectively {[}see Fig.
\ref{fig:Molecule-Binet}(b){]}. The sign of $L_{k}$ is conserved
on these trajectories, and depending on the enclosed pole (either
on positive or negative side of the axis), we denote the corresponding
sets by $T_{k}^{\pm}$. Although the Binet construction provides a
qualitative picture of $\mathbf{L}$ trajectories as seen from the
molecule fixed frame, it allows to deduce some valuable information
about the motion in the laboratory frame, as well. In case of a single
top, we can assume that the conserved vector of angular momentum,
$\boldsymbol{\ell}$ points along the laboratory $Z$ axis. For definiteness,
we assume $T_{c}$ trajectory, meaning that $\mathbf{L}$ moves on
a ``taco-shaped'' curve around one of the poles on the $c$ axis
{[}see Fig. \ref{fig:Molecule-Binet}(b){]}. This implies that in
the laboratory frame the $c$ axis ``precesses'' around $\boldsymbol{\ell}$
while the sign of the projection $\hat{\mathbf{c}}\cdot\boldsymbol{\hat{\ell}}=\hat{\mathbf{c}}\cdot\hat{\mathbf{Z}}$
remains unchanged, which means preferred orientation of the $c$ axis
in the course of time. Notably, even in case of an ensemble of asymmetric
tops, i.e. when initially angular momenta vectors point in various
directions, the permanent orientation is still possible. \textcolor{black}{In
that case} permanent orientation corresponds to the ensemble-averaged
quantities $\braket{\hat{\mathbf{a}}\cdot\hat{\mathbf{Z}}}$ or $\braket{\hat{\mathbf{c}}\cdot\hat{\mathbf{Z}}}$
having a constant sign. This may be achieved by, first, orienting
the averaged angular momentum vector $\braket{\boldsymbol{\ell}}$
(e.g. along $Z$ axis) and, second, breaking the symmetry between
$T_{k}^{+}$ vs $T_{k}^{-}$ trajectories.\\

\begin{figure}
\begin{centering}
\includegraphics[width=86mm]{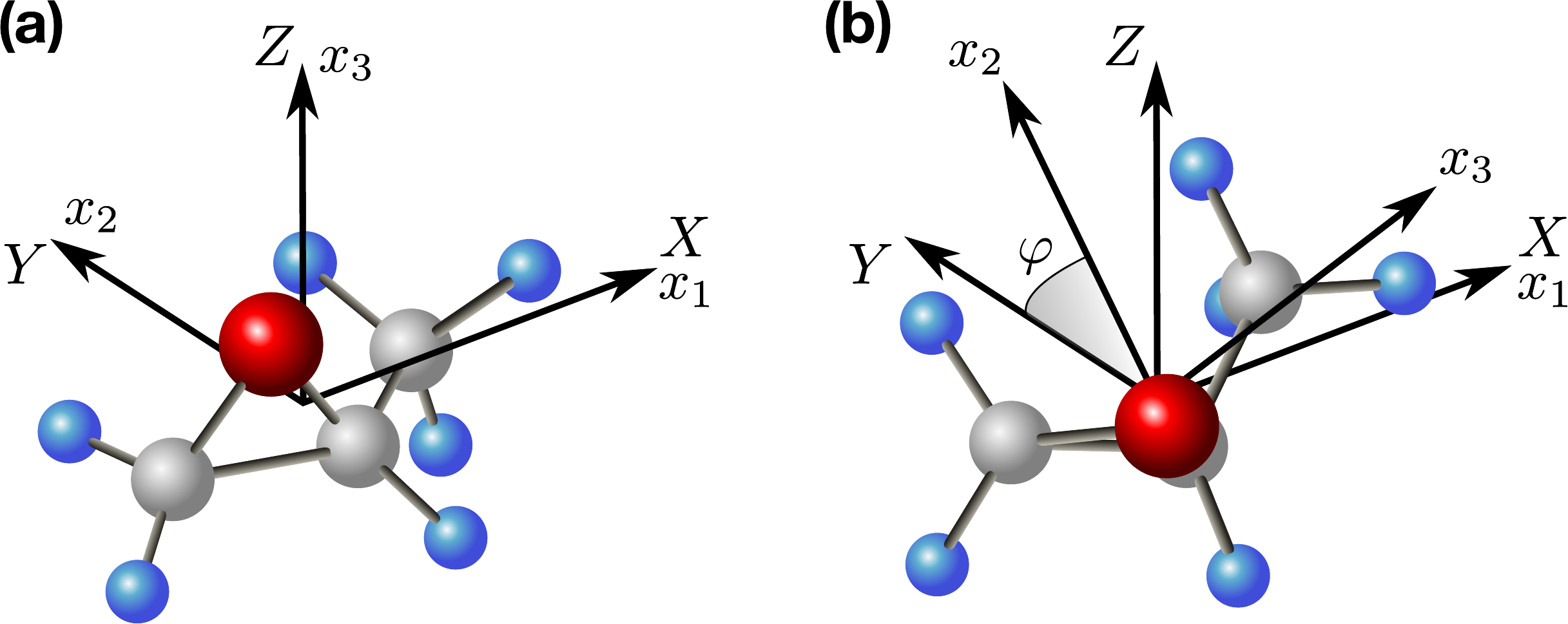}
\par\end{centering}
\caption{\textbf{(a)} The frame of polarizability principal axes $x_{1}$,
$x_{2}$ and $x_{3}$ ($\alpha_{33}<\alpha_{22}<\alpha_{11}$) coincides
with the laboratory fixed frame. \textbf{(b)} The molecule is rotated
(about $X$) by an angle $\varphi$, measured from the positive $Y$
axis to positive $x_{2}$ axis, lying in the $YZ$ plane. \label{fig:PPO-alpha-oriented-both}}
\end{figure}

\noindent \textbf{Particular case of twisted polarization. }Here we
consider a specific example of an ensemble of chiral molecules excited
by pair of delayed cross-polarized laser pulses \citep{Fleischer2009,Kitano2009,Khodorkovsky2011},
which constitutes the simplest implementation of a field with twisted
polarization. We show that such an excitation leads to the conditions
discussed above. Here, the first pulse is polarized along the laboratory
$X$ axis, while the polarization of the second one is in the $XY$
plane at $+\pi/4$ to the $X$ axis. The first pulse induces alignment
of the most polarizable molecular axis. For the sake of simplicity
of the qualitative analysis, we assume that after the first pulse
all the molecules are perfectly aligned along the $X$ axis, and are
stationary \citep{Gershnabel2018}. Initially, we consider only half
of all molecules, in which the most polarizable axis points along
$+X$ {[}see Fig. \ref{fig:PPO-alpha-oriented-both}(a){]}. Angle
$\varphi\in\left[0,2\pi\right)$ is the angle between the $Y$ axis
and the $x_{2}$ axis. lying in the $YZ$ plane {[}see Fig. \ref{fig:PPO-alpha-oriented-both}(b){]}.
The aligned molecules are uniformly distributed in $\varphi$. The
interaction potential, $U$ and torque, $\mathbf{T}$ induced by a
non-resonant optical field are given by
\[
U=-\frac{1}{2}\braket{\mathbf{d}_{\mathrm{ind}}\cdot\mathbf{E}}\qquad\mathbf{T}=\braket{\mathbf{d}_{\mathrm{ind}}\times\mathbf{E}},
\]

\noindent where the angle brackets denote time averaging over the
optical cycle, $\mathbf{d}_{\mathrm{ind}}=\boldsymbol{\upalpha}\mathbf{E}$
is the induced dipole, $\boldsymbol{\upalpha}$ is the polarizability
tensor, and $\mathbf{E}$ is the vector of the electric field. The
duration of the laser pulses is assumed to be short as compared to
the typical rotational periods of the chiral molecules, therefore
the effect of the second pulse is considered in the impulsive approximation,
$\Delta\mathbf{L}\propto\mathbf{T}$.

The second pulse (twisted with respect to the first one) induces unidirectional
rotation in the $XY$ plane resulting in orientation of $\braket{\boldsymbol{\ell}}$
along $Z$, which constitutes the first criterion for the long-lasting
orientation. In addition, there is an orienting torque acting on the
aligned most polarizable axis, $x_{1}$. To check whether the second
criterion for the long-lasting orientation is satisfied, we evaluate
the torques along the molecular $a$ and $c$ axes, $\tau_{k}$ ($k=a,c$)
\begin{equation}
\tau_{k}\propto C_{k}\sin\left(2\varphi\right)+D_{k}\sin\left(\varphi+\phi_{k}\right),\label{eq:principal-frame-torque}
\end{equation}

\noindent where $C_{k}=\left(\alpha_{33}-\alpha_{22}\right)R_{k1}/2$,
$D_{k}=\sqrt{A_{k}^{2}+B_{k}^{2}}$, $A_{k}=\left(\alpha_{11}-\alpha_{33}\right)R_{k2}$,
$B_{k}=\left(\alpha_{11}-\alpha_{22}\right)R_{k3}$, and $\cos\left(\phi_{k}\right)=A_{k}/D_{k}$.
Here, $R_{nm}$ are the elements of the orthogonal rotation matrix
relating the principal polarizability and inertia frames. (See Supplementary
Materials for details.) For chiral molecule, the two frames do not
align {[}see Fig. \ref{fig:Molecule-Binet}(a){]} and the off-diagonal
elements of $R$ differ from zero. In this case, the subdomains of
positive and negative $\tau_{k}$ in the interval $\varphi\in\left[0,2\pi\right)$
are not equal, which is precisely the required $T_{k}^{+}$ vs $T_{k}^{-}$
asymmetry. When the two frames do align (as in non-chiral molecules),
$R$ becomes diagonal resulting in torques $\tau_{a}=C_{a}\sin\left(2\varphi\right)$
and $\tau_{c}=A_{c}\sin\left(\varphi\right)$. In this case, the $T_{k}^{+}$
vs $T_{k}^{-}$ symmetry is preserved.

It may be shown that Eq. \ref{eq:principal-frame-torque} remains
the same for molecules with the most polarizable axis, $x_{1}$ pointing
along $-X$. For the complimentary enantiomer ((\emph{S})-PPO), signs
of some of the elements of $R_{nm}$ are reversed, resulting in the
opposite orientation direction.

\section{Numerical Simulations}

In this section, we present the results of numerical simulations of
the laser driven orientation dynamics of (\emph{R})-PPO molecule.
We consider two implementations of laser fields with twisted polarization,
a pair of delayed cross-polarized pulses \citep{Fleischer2009,Kitano2009,Khodorkovsky2011}
and an optical centrifuge \citep{Karczmarek1999,Villeneuve2000,Yuan2011,Korobenko2014,Korobenko2018}.
The behavior of an ensemble of $N\gg1$ molecules is investigated
using both classical and quantum mechanical tools. The chiral molecule
is modeled as a rigid asymmetric top having anisotropic polarizablity
and a dipole moment. Supplementary Table 1 summarizes the molecular
properties used.

For the classical simulations, the behavior of a thermal ensemble
was simulated using the Monte Carlo approach. Our numerical scheme
relies on solving the Euler equations for angular velocities and parametrizing
the rotations by quaternions \citep{Art-of-Molecular-Simulation}.
The details of the scheme may be found in \citep{Tutunnikov2018}.
Before the excitation, the molecules are isotropically distributed,
and the angular velocities are assigned according to the Boltzmann
distribution for a given temperature.

For the quantum simulations, we use the symmetric top wavefunctions
$\ket{JKM}$ as a basis set \citep{Zare}. Here $J$ is the total
angular momentum (in Section \ref{sec:Classical-Analysis} we used
the label $L$ for the modulus of the classical angular momentum),
$K$ is its projection on the molecule-fixed $c$ axis, and $M$ is
its projection onto the laboratory fixed $Z$ axis. In this basis,
the matrix representing the kinetic energy Hamiltonian has a tridiagonal
form, in which the states of different $K$'s are coupled with $K\pm2$
states (see Sup. Eq. 8). The interaction potential is expressed in
terms of Wigner D-matrices and its matrix elements are evaluated in
the $\ket{JKM}$ basis. The explicit expressions used may be found
in the Supplementary Material. Then, a unitary transformation is applied
transforming the matrices to the asymmetric-top basis, in which the
kinetic energy Hamiltonian becomes diagonal. All the thermally populated
(for a given temperature) eigenstates of the chiral molecule are propagated
in time separately, and thermal averaging is performed to obtain the
observables of interest. The explicit expressions for matrix elements
of the observables are presented in the Supplementary Material, as
well.\\

\noindent \textbf{Double pulse excitation.} Here we consider excitation
of an ensemble of (\emph{R})-PPO molecules by a pair of delayed cross-polarized
laser pulses. The first pulse is polarized along the laboratory $X$
axis, while the polarization of the second one is in the $XY$ plane
at $+\pi/4$ angle to the $X$ axis. The electric field of the pulses
is given by $\boldsymbol{\mathcal{E}}_{i}\left(t\right)=\mathcal{E}_{0}\exp\left[-2\ln2\left(\left(t-t_{j}\right)/\mathrm{FWHM}\right)^{2}\right]\cos\left(\omega t\right)\mathbf{e}_{i}$,
where $\mathcal{E}_{0}$ is the maximal amplitude of the electric
field of the pulse, FWHM is the full width at half maximum of the
intensity profile, $\omega$ is the carrier frequency of the pulse,
and $\mathbf{e}_{i}$ is a unit vector along the polarization direction
($i=1,2$). The maximal amplitude of the electric field is related
to the peak intensity by $I_{0}=\varepsilon_{0}cE_{0}^{2}/2$, where
$\varepsilon_{0}$ is the permittivity of vacuum and $c$ is the speed
of light in vacuum.

\noindent The simulated classical results span the range of temperatures
$T\in\left[10,100\right]$ K with a step of 10 K. The first pulse
leads to the alignment of the most polarizable molecular axis along
$X$, and the second pulse is applied when this alignment reaches
the maximal value. Since this value as well as the time required to
reach it are temperature dependent, the moment of application of the
second pulse was optimized for each temperature.

\begin{figure}
\begin{centering}
\includegraphics[width=86mm]{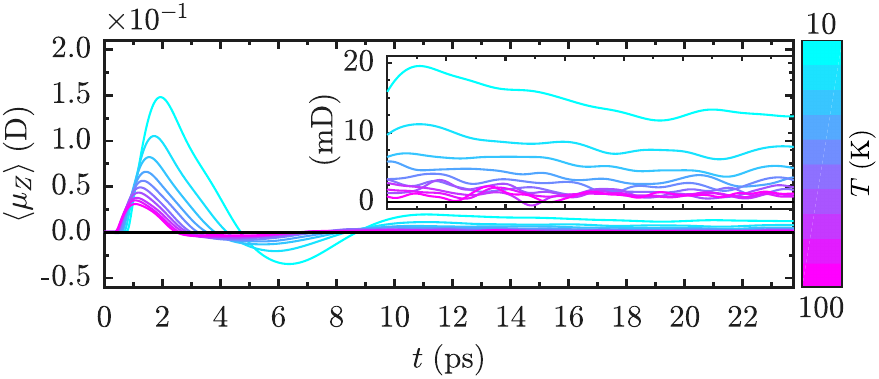}
\par\end{centering}
\caption{Classical ensemble averaged projection of the molecular dipole on
$Z$ axis, $\braket{\mu_{Z}}\left(t\right)$ for the case of (\emph{R})-PPO
molecules excited by a pair of cross-polarized pulses. Here $N=6\times10^{6}$.
The curves span the range of temperatures $T\in\left[10,100\right]$
K, in steps of 10 K. Parameters of the pulses are: $I_{0}=0.5\times10^{14}\;\mathrm{W/cm^{2}}$,
$\mathrm{FWHM=0.10}$ ps. First pulse is centered at $t=0$, while
the delay of the second pulse is optimized for each $T$ (see text).
Horizontal line is set at $\braket{\mu_{Z}}=0$. \textcolor{black}{The
inset is an amplified version of the figure.} \label{fig:PPO-double-pulse}}
\end{figure}

Figure \ref{fig:PPO-double-pulse} shows the ensemble-averaged projection
of the molecular dipole on the $Z$ axis as a function of time, $\braket{\mu}\left(t\right)$
resulting from excitation by both pulses. As may be seen, the second
pulse results in a sharp transient orientation of the ensemble-averaged
dipole reaching an approximately constant value in the long term.
It is important to emphasize that the orientation direction is perpendicular
to the plane of twisting and its sign depends on both the sense of
twisting and the handedness of the molecule. The sign of orientation
is opposite for the second enantiomer (\emph{S})-PPO. The classical
mechanism explaining the induced enantioselective transient orientation
is described elsewhere \citep{Gershnabel2018,Tutunnikov2018}. The
maximal amplitudes of both the transient and the persistent dipole
are temperature dependent and become lower with increasing the temperature.
The long-lasting dipole signals shown in Figure \ref{fig:PPO-double-pulse}
exhibit small-amplitude beats about the asymptotically constant values.
These beats stem from the statistical nature of the simulations, and
they are inversely proportional to $\sqrt{N}$.

\begin{figure}
\begin{centering}
\includegraphics[width=86mm]{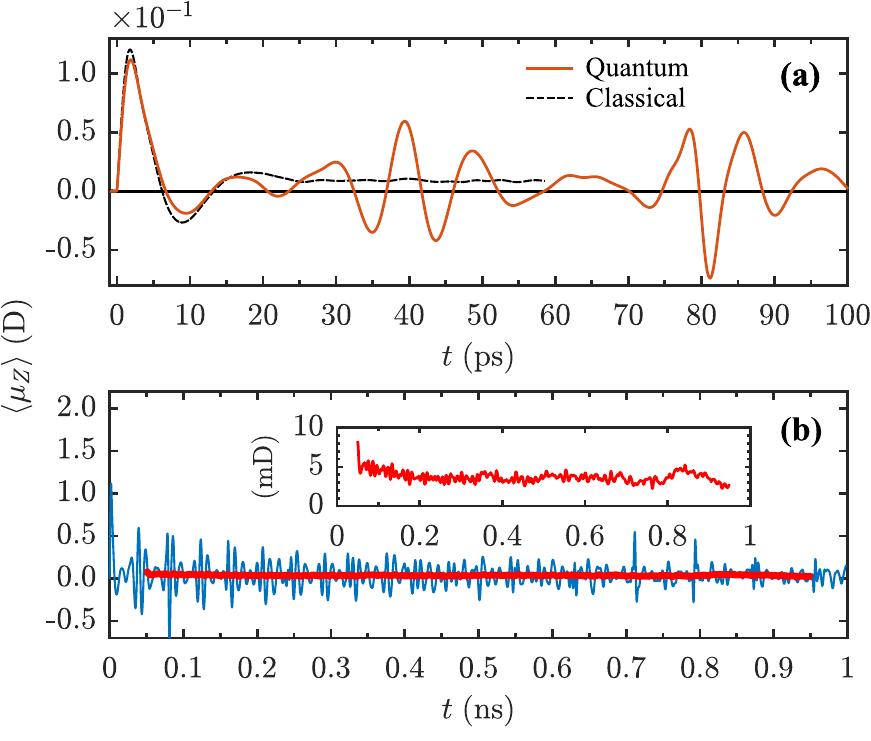}
\par\end{centering}
\caption{Quantum expectation value of projection of the molecular dipole on
$Z$ axis, $\braket{\mu_{Z}}\left(t\right)$ for the case of (\emph{R})-PPO
molecules excited by a pair of cross-polarized pulses. The temperature
is set at $T=5$ K. Parameters of the pulses are: $I_{0}=1.43\times10^{14}\;\mathrm{W/cm^{2}}$,
$\mathrm{FWHM=20}$ fs. The first pulse is centered at $t=0$, the
delay of the second pulse is 1.2 ps. \textbf{(a)} The short-time dipole
signal (solid) and its classical counterpart (dashed). \textbf{(b)}
Long-time dynamics. Solid red - 100 ps time-average, $\overline{\braket{\mu_{Z}}\left(t\right)}=\frac{1}{\Delta t}\int_{t-\Delta t/2}^{t+\Delta t/2}\braket{\mu_{Z}}\left(t^{\prime}\right)\mathrm{d}t^{\prime}$.
Inset is a magnified portion of the figure. \label{fig:PPO-double-pulse-QC}}
\end{figure}

To access the influence of the quantum effects on the permanent orientation,
we carried out fully quantum mechanical simulations of the dynamics
of (\emph{R})-PPO molecules kicked by the two pulses. The pulse envelope
used in the quantum simulations, is given by Sup. Eq. 6. The difference
between this envelope and the Gaussian one used in the classical simulation
is insignificant, as the pulses are short and the difference in their
integrals is negligible ($<1\%$). The rotational temperature was
set to $T=5$ K. The delay of the second pulse was adjusted to the
moment of maximal alignment of the most polarizable axis towards the
$X$ axis, as measured using the alignment factor (see Sup. Eq. 23).
Depending on the initial conditions, the pulses excite the molecules
to the states with $J$ up to $20\hbar$, with the mean value of about
$8\hbar$, which is above the average thermal value of $3.5\hbar$.
The short-time dipole signal (see Sup. Eq. 18) is shown in Fig. \ref{fig:PPO-double-pulse-QC}(a)
with its classical counterpart for comparison. Shortly after the pulse,
the polarization peaks at about $0.12~\mathrm{D}$. In contrast to
the classically predicted permanent steady-state dipole signal, the
quantum curve exhibits beats, spaced by $\approx40$ ps (the first
few). These quantum beats are a well-known phenomenon for quantum
rotors \citep{Felker1992}. However, the long-term time-averaged signal
differs from zero, see Figure \ref{fig:PPO-double-pulse-QC}(b). The
inset of the figure shows the time average, as defined in the figure's
caption. \\

\begin{figure}
\begin{centering}
\includegraphics[width=86mm]{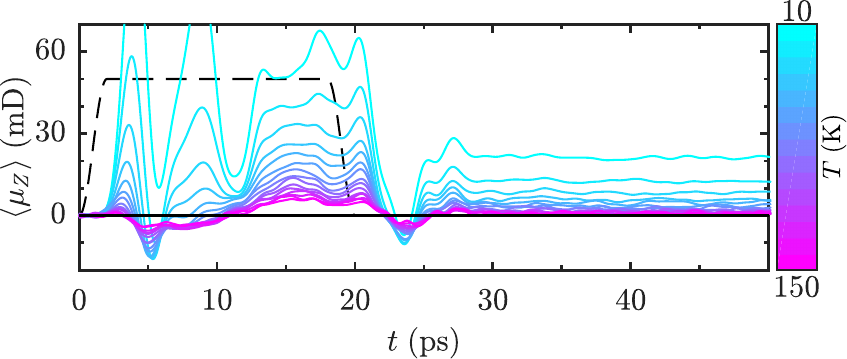}
\par\end{centering}
\caption{Classical ensemble averaged projection of the molecular dipole on
the $Z$ axis, $\braket{\mu_{Z}}\left(t\right)$ for the case of (\emph{R})-PPO
molecules excited by the optical centrifuge. The curves span the range
of temperatures $T\in\left[10,150\right]$ K, in steps of 10 K. Here
$N=9\times10^{6}$. Parameters of the pulse are: $I_{0}=5\times10^{12}\;\mathrm{W/cm^{2}}$,
$\beta=0.065\;\mathrm{ps^{-2}}$ and $t_{\mathrm{off}}=t_{\mathrm{on}}=2$
ps and $t_{\mathrm{p}}=20$ ps. The dashed curve represents the scaled
$\varepsilon^{2}(t)$, see Eqs. \ref{eq:centrifuge-field} and Sup.
Eq. 7. The horizontal line is set at $\braket{\mu_{Z}}=0$. \label{fig:PPO-centrifuge}}
\end{figure}

\begin{figure}
\begin{centering}
\includegraphics[width=86mm]{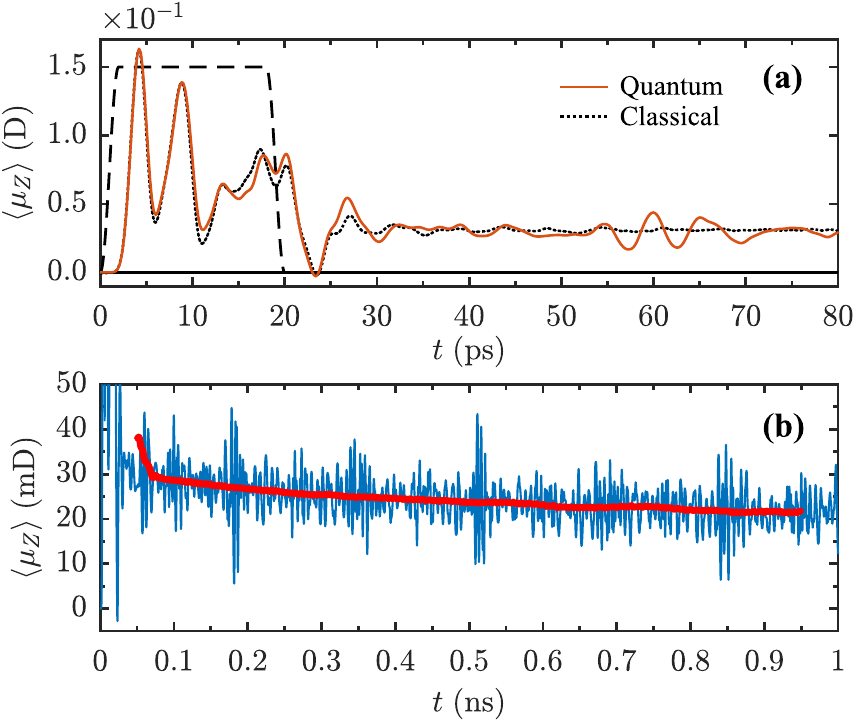}
\par\end{centering}
\caption{Quantum expectation value of the projection of the molecular dipole
on the $Z$ axis, $\braket{\mu_{Z}}\left(t\right)$ for the case of
(\emph{R})-PPO molecule excited by the optical centrifuge pulse. Parameters
are similar to those of Fig. \ref{fig:PPO-centrifuge}. \textbf{(a)}
The short-time dipole signal (solid) and its classical counterpart
(dotted). The dashed curve represents the scaled $\varepsilon^{2}(t)$,
see Eqs. \ref{eq:centrifuge-field} and Sup. Eq. 7. \textbf{(b)} Long-time
dynamics. Solid red - 100 ps time-average, $\overline{\braket{\mu_{Z}}\left(t\right)}=\frac{1}{\Delta t}\int_{t-\Delta t/2}^{t+\Delta t/2}\braket{\mu_{Z}}\left(t^{\prime}\right)\mathrm{d}t^{\prime}$\label{fig:PPO-centrifuge-QC}}
\end{figure}

\noindent \textbf{Optical centrifuge excitation.} Optical centrifuge
is a laser pulse, whose linear polarization undergoes an accelerated
rotation around its propagation direction \citep{Karczmarek1999,Villeneuve2000,Yuan2011,Korobenko2014,Korobenko2018}.
We model the electric field of such a pulse by
\begin{equation}
\mathbf{\boldsymbol{\mathcal{E}}}=\varepsilon(t)\left[\cos\left(\beta t^{2}\right)\mathbf{e}_{X}+\sin\left(\beta t^{2}\right)\mathbf{e}_{Y}\right]\cos\left(\omega t\right),\label{eq:centrifuge-field}
\end{equation}
where $\beta$ is the angular acceleration and $\varepsilon(t)$ is
the envelope (dashed curve, Fig. \ref{fig:PPO-centrifuge} and \ref{fig:PPO-centrifuge-QC}),
for explicit expression see Sup. Eq. 7. As in the previous case, we
start with the results of the classical simulations. Figure \ref{fig:PPO-centrifuge}
shows the ensemble-averaged projection of the molecular dipole on
the $Z$ axis as a function of time, $\braket{\mu_{Z}}\left(t\right)$
resulting from excitation by the optical centrifuge. The temperature
changes in steps of 10 K in the range of $T\in\left[10,150\right]$
K.
Akin to the double pulse scheme, excitation by the optical centrifuge
results in enantioselective dipole signal. The signal drops down when
the centrifuge is switched off after 20 ps. However, the orientation
signal does not vanish completely, but levels off at a non-zero value.
Although the amplitude of the transient signal is higher in the case
of the double pulse excitation, the amplitudes of the long-time signals
are higher in this scenario, making the optical centrifuge more efficient
in inducing the permanent dipole orientation. The amplitudes of the
signals, both during the driven and field-free dynamics periods are
temperature dependent and decrease with temperature.

To investigate the long-lasting dynamics of the induced dipole moment,
we carried out a fully quantum mechanical simulation of the centrifuge-driven
(\emph{R})-PPO rotational dynamics. Centrifuge pulse had the same
parameters as shown in the caption to Fig. (5), while the rotational
temperature was set to $T=5$ K. Such a pulse excites angular momenta
of up to $40\hbar$, with the mean excitation of about $30\hbar$,
which is higher as compared to the previous double-pulse excitation
example, and significantly above the average thermal value of $3.5\hbar$.
Higher angular momentum results in a better quantum-classical agreement,
as compared to the double pulse excitation. The short-time dipole
signal is shown in Fig. \ref{fig:PPO-centrifuge-QC}(a). During the
pulse, the polarisation peaks at about $0.15~\mathrm{D}$, and it
stays practically constant at the level of $\approx0.03~\mathrm{D}$
after the end of the pulse, with a small-amplitude beating structure
present at long times. For comparison, the classical signal obtained
under the same conditions is plotted and the correspondence with the
quantum results is very good up to $t\approx60$ ps. On the nanosecond
time-scale {[}Fig. \ref{fig:PPO-centrifuge-QC}(b){]}, one can observe
multiple beats, however the coarse-grained time-averaged quantum signal
remains positive, and almost constant as the classically predicted
one. \\

\noindent \textbf{Dynamical Tunneling.} Due to the coupling of different
$K$-states by the rotational Hamiltonian (see Sup. Eq. 8), the quantum
mechanical rigid asymmetric top does not have eigenstates with the
angular momentum being oriented within the molecular frame. As a consequence,
any state that is internally oriented at some point in time can not
be an eigenstate and will oscillate between being oriented and anti-oriented,
an effect known as dynamical tunneling \citep{Keshavamurthy2011}.
In other words, a clockwise internal rotation would eventually become
counter-clockwise, and vice versa. This leads to a breakdown of quantum-classical
analogy, as the classical model introduced in Sec. \ref{sec:Classical-Analysis}
assumes a strict separation of clockwise and counter-clockwise rotation
(no interchange between $T_{k}^{+}$ and $T_{k}^{-}$). Therefore,
one would expect no permanent orientation after turn-off of all external
fields for a quantum mechanical chiral rotor.

Yet, in spite of the dynamical tunneling, our simulations show a significant
long-time orientation of the quantum rotors, equal in magnitude to
the classical model. The reason is that the dynamical tunneling time
grows very fast with the angular momentum \citep{Harter1984}, being
larger the more the angular momentum is (anti-)oriented along the
$a$ or $c$ axis. Indeed, already for $J=10\hbar$, the tunneling
time may exceed microseconds \citep{Harter1984}, explaining why orientation
was observed in the quantum simulations on a nanosecond time-scale.
As a matter of fact, the slight decrease of the base line visible
in Fig. \ref{fig:PPO-centrifuge-QC}b (thick red line) is possibly
a signature of the dynamical tunneling.

At this point one should also note that whilst dynamical tunneling
would be of greater importance on longer time-scales than the ones
presented here, other effects (neglected here), like rovibrational
couplings, hyperfine structure \citep{Thomas2018} and intermolecular
collisions are likely of higher importance and would probably bury
any signs of the dynamical tunneling in an actual experiment.\\

\begin{figure}
\begin{centering}
\includegraphics[width=86mm]{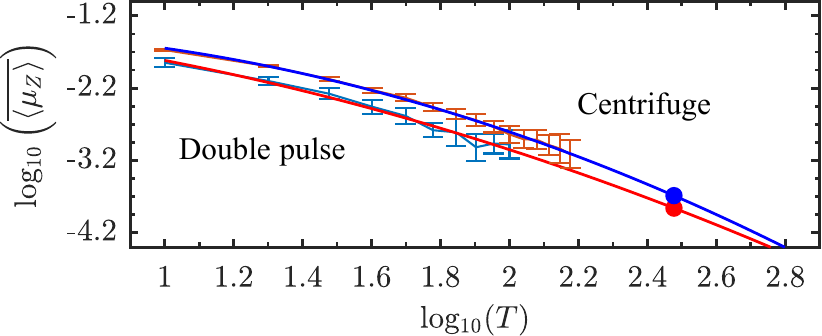}
\par\end{centering}
\caption{Logarithmic scale plot of the average constant component of $\braket{\mu_{Z}}$,
$\overline{\braket{\mu_{Z}}}=\frac{1}{t_{\mathrm{max}}-\tau}\int_{\tau}^{t_{\mathrm{max}}}\braket{\mu_{Z}}\left(t^{\prime}\right)\mathrm{d}t^{\prime}$
as a function of temperature. For the double pulse excitation $\tau=12$
ps, while for the centrifuge $\tau=30$ ps (Figs. \ref{fig:PPO-double-pulse}
and \ref{fig:PPO-centrifuge}). The error bars are obtained using
the standard error propagation formula \citep{Ku1966}. Solid lines
are the fit function, $f\left(x\right)=ax^{2}+bx+c$. The parameters
triplets $\left(a,b,c\right)$ are $\left(-0.31,-0.30,-1.20\right)$
and $\left(-0.47,0.24,-1.42\right)$. The bold dots correspond to
$T=300$ K and their coordinates are $\left(2.48,-3.69\right)$ and
$\left(2.48,-3.86\right)$ for the centrifuge and double pulse, respectively.
\label{fig:Log-log-plot-PPO}}
\end{figure}
\noindent \textbf{Extrapolation to room temperature. }Due to the statistical
nature of our classical simulations on one hand and basis size limitations
of the quantum simulation on the other, direct simulations at room
temperature are highly numerically demanding, therefore we resort
to extrapolation. Figure \ref{fig:Log-log-plot-PPO} shows the average
long term value of the classically calculated $\braket{\mu_{Z}}$,
denoted by $\overline{\braket{\mu_{Z}}}$ as a function of temperature
on the double logarithmic scale. The error bars for $\log_{10}(\overline{\braket{\mu_{Z}}})$
are obtained by the standard error propagation formula \citep{Ku1966}
based on the variance of $\braket{\mu_{Z}}$. The points are fit by
a polynomial and each point is assigned a weight equal to inverse
of its variance. The fit allows to extrapolate the results to temperatures
beyond those simulated. The predicted value of the permanent dipole
at room temperature is $\approx10^{-3.9}\;\mathrm{D}$ and $\approx10^{-3.7}\;\mathrm{D}$,
for the double pulse and centrifuge excitation schemes, respectively
(see Fig. \ref{fig:Log-log-plot-PPO}). Because of the relatively
high vapor pressure of propylene oxide at room temperature \citep{vapor},
even such modest values of the mean molecular dipole moment result
in a sizable macroscopic polarization in the gas sample, and cause
a collective electric field of about 0.05 V/m in the focal region
of a laser beam focused to a spot of $r_{0}\sim10\;\mathrm{\mu m}$
and having the Rayleigh range of about $z_{\mathrm{R}}\sim1\;\mathrm{mm}$.
The magnitude and longevity of this field (and the corresponding light-induced
voltage) favor its detection using high-speed nanosecond electronics.

The appearance of long-time orientation in a gas sample can be detected
by means of Coulomb explosion \citep{Pitzer2013,Herwig2013,Christensen2015,Pitzer2018,Fehre2019,Milner2019}
and nonlinear optics through high harmonic generation of various orders
\citep{Kamta_2005,Frumker2012,Frumker2012b,Kraus2014}, especially
by means of the second harmonic generation. Another approach to detecting
the induced orientation is by measuring THz emission due to free induction
decay of coherently oscillating molecular dipoles \citep{Harde1991,Fleischer2011,Babilotte2016}.

\section{Conclusions}

Summarizing, we have demonstrated long-lived orientation of chiral
molecules by laser fields with twisted polarization using both classical
mechanics and fully quantum-mechanical treatments. The problem was
analyzed using propylene oxide molecules, as an example, and considering
two different implementations of the twisted optical field: (i) a
pair of delayed cross-polarized laser pulses, and (ii) an optical
centrifuge. We found very good agreement between the classical and
quantum approaches over a wide range of experimentally relevant parameters,
and over an extended time range. Significantly, we demonstrated the
novel chiral phenomenon of persistent molecular orientation lasting
long after the end of the exciting pulses. This counterintuitive effect
was first conjectured in our previous papers \citep{Gershnabel2018,Tutunnikov2018}
utilizing a classical approach, but its validity is now verified on
a much longer time scale (several orders of magnitude longer) where
the use of quantum treatment is unavoidable. The sign of the oriented
dipole moment depends on both the sense of polarization twisting and
the handedness of the molecule. This long-lasting orientation provides
new modalities for detecting molecular chirality with the help of
optical harmonics generation, or direct high-speed measurements of
electric fields caused by laser-induced macroscopic polarization of
the gas. Moreover, it is well known that optical pre-orientation of
molecules affects their deflection by inhomogeneous fields (see e.g.
\citep{Gershnabel2011a,Gershnabel2011b,Yachmenev2019}, references
therein, and recent reviews \citep{Fleischer2012,Lemeshko2013,Chang2015}).
Enantioselective orientation effects considered in this paper may
open up new avenues for separation of molecular enantiomers.

We note that ongoing experiments in Milner\textquoteright s group
show initial signals of the long-time enantioselective orientation,
supporting our theoretical predictions. A report on these results
and the corresponding theoretical analysis is in preparation \citep{LongChiralExp_tobepub}.

\section*{Acknowledgments}

The authors appreciate useful discussions with A. A. Milner and V.
Milner. This work was supported by the Israel Science Foundation (Grant
No. 746/15), the ISF-NSFC joint research program (Grant No. 2520/17),
\textcolor{black}{and by Natural Sciences and Engineering Research
Council of Canada grant to PB.} IA acknowledges support as the Patricia
Elman Bildner Professorial Chair, and thanks the UBC Department of
Physics \& Astronomy for hospitality extended to him during his sabbatical
stay. This research was made possible in part by the historic generosity
of the Harold Perlman Family.

\section*{References}

\bibliographystyle{unsrt}

\end{document}